\documentclass[twocolumn,3p]{elsarticle}
\usepackage{amscd, graphicx, fullpage, epstopdf, appendix, hyperref, color, setspace, tensor, url, verbatim, float, natbib, mathrsfs, caption, subcaption, amsmath,amssymb}
\title{}
\date{}
\newcommand{\beq}{\begin{equation}}
\newcommand{\ee}{\end{equation}}

\newcommand{\ket}[1]{\vert #1\rangle}

\newcommand{\evalue}[1]{\left\langle #1 \right\rangle}

\newcommand{\hide}[1]{}

\newcommand{\eq}[1]{Eq.\,(\ref{#1})}
\newcommand{\eqs}[1]{Eqs.\,(\ref{#1})}
\newcommand{\noeq}[1]{(\ref{#1})}
\newcommand{\fig}[1]{Fig.\,\ref{#1}}

\biboptions{sort&compress}

\begin{document}
\title{Limits and Possibilities of Refractive Index in Atomic Systems}

\author[uconn]{Robert A. McCutcheon\corref{corr}}
\cortext[corr]{Corresponding author. \\ \textit{E-mail address:} robert.mccutcheon@uconn.edu \\ \textit{Postal address:} Department of Physics \\
University of Connecticut Unit 3046 \\
196 Auditorium Road, Storrs, CT 06269-3046 \\
\textit{Phone number:} (941) 993-7860}
\author[uconn,harvard]{Susanne F. Yelin}

\address[uconn]{Department of Physics, University of Connecticut, Storrs, CT 06269}
\address[harvard]{Department of Physics, Harvard University, Cambridge, MA 02138}

\date{July 27, 2021}

\begin{abstract}
Manipulation of the refractive index has been of growing interest lately. We consider parameters and possibilities of enhancing the absolute-value limit of the linear index in coherent atomic systems. Starting with a review of how two-level transitions -- without and  with added coherence effects -- do not realistically allow for a significantly enhanced index at fixed atomic density, we discuss a possible way around this using wave mixing. This confirms that the only parameters, besides the medium optical depth/density, that can effectively change the value of the attainable index are the frequencies of the involved transitions. 
\end{abstract}

\begin{keyword}
Enhanced refractive index, linear refractive index, limitations on refractive index, wave mixing
\end{keyword}
\maketitle

\section{Introduction}\label{Intro}
The values of the refractive index most commonly observed in nature are positive and on the order of 1. Finding materials with an unusual refractive index -- very high, zero, or negative -- has long been a subject of interest; of course, to be of practical use, it should be accompanied by low absorption. It was proposed that quantum interference effects could be used to create an enhanced index of refraction with no absorption in $\Lambda$-type atoms \cite{Scully,Dowling-Bowden}. However, it was later noted that at the densities required for such methods, quantum corrections such as cooperative effects must be considered \cite{Y-F, F-Y,Fleischhauer}, which can dramatically diminish the promise of the original proposals. For example, radiation trapping effectively decreases the effects of coherence. Moreover, a new article \cite{chang21} reveals that densities that are effective for increasing the index of refraction are severely limited due to disorder effects even if the real densities are very high and nonlinear effects are disregarded.

\begin{figure}[h!tb]
\centering
\includegraphics[width=\linewidth]{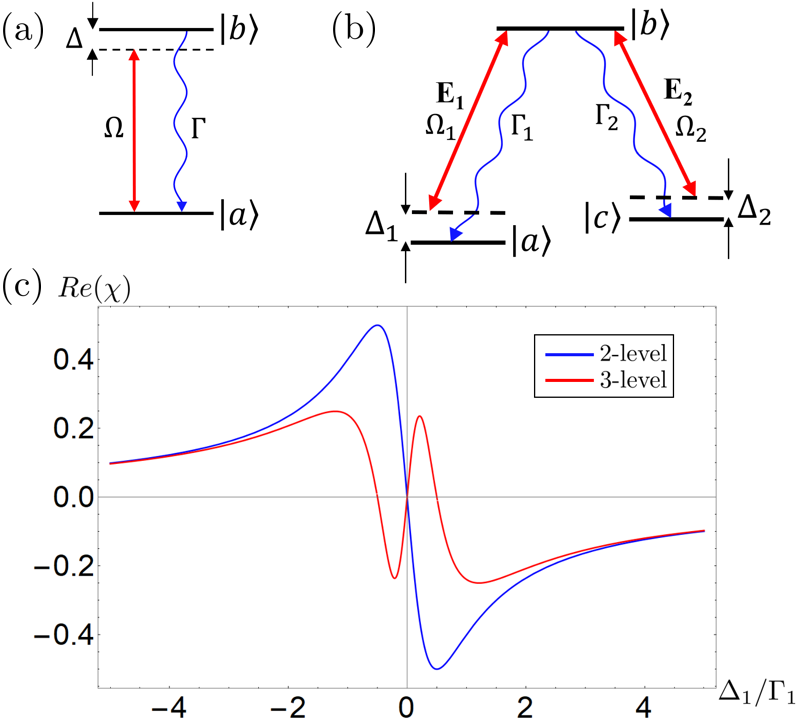}
\caption{ Two- and three-level systems do not allow to increase the maximum attainable index of refraction, keeping the density fixed. (a) Two-level system with characteristic frequency $\omega$, detuning $\Delta$, Rabi frequency $\Omega$, and population decay rate $\Gamma$, coupled to the external field $\mathbf{E}$. (b) Three-level system. $\mathbf{E_i}$ couples to transition $i=1,2$ with characteristic frequency $\omega_i$, dipole operator $\hat{\mathbf{d}}_i$, decay rate $\Gamma_i$, Rabi frequency $\Omega_i$, and detuning $\Delta_i$. (c) Two-level and three-level system susceptibilities]{Real part of susceptibility for the two-level system (blue curve) and three-level system (red curve). $N=10^{12}$ cm\textsuperscript{-3}, $\Gamma\equiv\Gamma_1$, $\Gamma_2=\Gamma_1$, $\Omega\equiv\Omega_1=\Gamma_1$, $\Omega_2=\Gamma_1$, $\Delta_2=0$, $\gamma_0=\Gamma_1/100$, $\omega\equiv \omega_1=2\pi\times 10^{14}$ s\textsuperscript{-1}.}}
\label{2_level_vs_3_level}
\end{figure}

Since then, there have been increasing efforts to modify the index of refraction of a system or to attain a particular value through the design of metamaterials \cite{meta1,meta2,meta3} or control by external fields \cite{Kastel,density1,fieldcontrol,Marina}. Unusual values for the index have been suggested for use in various applications; for example, a negative index could be used to implement cloaking \cite{cloaking,cloaking2,cloaking3,cloaking4} or to create a ``perfect lens'' with infinite resolution \cite{ni,lens}, while a medium with a large index decreases the wavelength of light traveling through it, which could be useful in optical imaging \cite{nanooptics}.

Let us briefly clarify what we mean by ``high index of refraction'' in the rest of this article: the real part of the atomic electric susceptibility -- and thus the index of refraction -- are strongly frequency-dependent, especially close to resonance. Thus, the first necessity of finding a very high positive or negative susceptibility is to ensure that the maxima and minima of the susceptibility spectrum are high. This is what is studied in this article. Selecting optimal detuning from resonance and making sure that the respective absorption value is close to zero would be the second step. This has been treated, however, in depth in the articles cited above.

There are many parameters that can be used to control an index of refraction in some medium. In theoretical approaches focusing on the use of external fields to change the optical response of a system, these parameters can simply be changed in order to modify the index of refraction. However, there are limits to what values of index are possible and we start this article by briefly reviewing that changing the (effective) density of the atomic medium seems to be the only way to change the magnitude of the index apart from its frequency. The main part of the article is devoted to show how wave mixing can be used to change the effective frequency and thus the index of refraction without changing the density of the gas.

The index of refraction of a medium is defined by $n=\sqrt{\epsilon\mu}$, with relative permittivity $\epsilon=1+\chi_e$ and relative permeability $\mu=1+\chi_m$ \cite{text1,grynberg10,jackson98}. The electric susceptibility $\chi_e$ and magnetic susceptibility $\chi_m$ can be complex, resulting in a complex index $n$, for which the real part is the ``conventional" refractive index which represents the amount of refraction, while the imaginary part represents the amount of absorption or gain. In this paper we assume that $\mu\approx 1$, since magnetic coupling is typically weaker by a factor of $\alpha^2\approx 1/137^2$ for transitions in the visible spectrum, so that $n\approx\sqrt{1+\chi_e}$ (henceforth we drop the subscript ``$e$"). 

Although the real part of $n$ depends on a complex $\chi$, in practice, ultimately one would want zero or minimal absorption, where the imaginary part of $\chi$ is negligible. We are here interested only in the best-case scenarios for enhancing the real part of $n$, so we will focus on the real part of $\chi$ in Section \ref{limits section}.

We suppose that electric fields of the form $\mathbf{E}(z,t)=\frac{1}{2}\hat{\mathbf{\epsilon}}\mathcal{E}(z)e^{i(kz-\nu t)} + \text{c.c.}$ each couple to one atomic transition with dipole operator $\hat{\mathbf{d}}$ in an ensemble of atoms with number density $N$, which do not interact with each other. With linear dispersion, the polarization due to such an electric field has the form $\mathbf{P}(z,t)=\frac{1}{2}\hat{\mathbf{\epsilon}}\mathcal{P}(z)e^{i(kz-\nu t)} + \text{c.c.}$, and $\mathcal{P}(z)=\epsilon_0 \chi \mathcal{E}(z)$ if the polarization depends on only one field. The susceptibility can be calculated by equating this with $\mathcal{P}(z)=2Nd^{*} \rho_{ij}$, since $\mathbf{P}(z,t)=N\evalue{\hat{\mathbf{d}}}(z,t)$, and solving the optical Bloch equations for the relevant $\rho_{ij}$ in the steady state \cite{text1, grynberg10, jackson98}. 

In the following sections, we consider the index experienced by a field on the $\ket{a}-\ket{b}$ transition in three cases. In the simplest case of an index due to a single transition between two atomic levels, there are significant limitations to the index that can be experienced by a field propagating through an ensemble of such atoms. Even with coherence effects introduced by an additional transition, this result is not improved. These cases are reviewed in Sec. \ref{limits section} in the context of enhanced index to find baseline values for the index and point out their limitations. In Sec. \ref{4L section}, wave mixing is considered as a new effect that could possibly be used to go beyond these limitations and obtain an enhanced index. We show how the direct frequency dependence that occurs gives some more flexibility in enhancing the index, and calculate how the index experienced by one field can be affected by the frequency of another field, which is the main result of this paper. However, we also find that there are important limitations to this method as well.

\begin{centering}
\section{Limits of linear index of refraction}\label{limits section}
\end{centering}

\subsection{Basic case -- two levels}\label{2L section}

We start with the most basic result which comes from a two-level atom driven by an external field $\mathbf{E}$ (shown in \fig{2_level_vs_3_level}a) in order to find baseline values for $\chi$ and therefore $n$ to which other values can be compared, while considering the natural limitations in attaining them.

The standard result for the susceptibility is \cite{text1, Loudon}
\begin{equation}
  \chi = N\frac{3\lambda^3\Gamma}{4\pi^2}\frac{i\Gamma-2\Delta}{\Gamma^2+4\Delta^2+2|\Omega|^2},  
\end{equation}
where $\lambda$ is the wavelength of the incident light, and the detuning $\Delta$ is typically much smaller than the atomic resonance or field frequency.
For a given $\lambda$, it is evident that the density $N$ is the only parameter that could potentially be varied in order to modify the absolute value of the susceptibility by an appreciable amount. As an example, we will estimate the density needed to give a maximum for $Re(\chi)$ on the order of $1$. This is
\begin{equation}
{\rm Re}(\chi)_{max} \;\approx\; \frac{3}{8\pi^2}N\lambda^3 \;=\; \mathcal{O}(1).    
\end{equation}
For optical wavelengths, this would require a density on the order of $10^{14}$ atoms per cm\textsuperscript{3}. The susceptibility could be theoretically increased by increasing the density, but values greater than $10^{14}$ cm\textsuperscript{-3} are not realistically possible \cite{Marina,density4}. At such high densities, collisions and nonlinear effects become dominant compared to the linear contribution to the dispersion, so this simple model is no longer valid \cite{chang21,density1,Thommen,density2,density3,javanainen16,jennewein18,pellegrino14,boyd92}.

The real part of $\chi$ is shown in \fig{2_level_vs_3_level}c (with $\Delta/\Gamma \equiv \Delta_1/\Gamma_1$), along with the corresponding three-level results which will be discussed in the next section.

\subsection{Coherence effects -- three levels} \label{3L section}
In the interest of finding a way to modify the susceptibility and refractive index beyond the result of Sec. \ref{2L section}, we consider coherent modification by including an additional level, as done in \cite{Scully, Dowling-Bowden, text1}, and discuss the limitations on the linear susceptibility. 

We use the $\Lambda$-system of \fig{2_level_vs_3_level}b. We also include a typically small decoherence rate $\gamma_0$ between levels $\ket{a}$ and $\ket{c}$, but find that this does not significantly affect the results. The coherence related to the polarization due to the transition of interest is $\rho_{ba}$:
\begin{equation}
\rho_{ba}=\frac{\Omega_{1}(\rho_{bb}-\rho_{aa})-\Omega_{2}\rho_{ca}}{i(\Gamma_{1}+\Gamma_{2})+2\Delta_1}.
\label{rho31}
\end{equation}
The main contribution from the third level is seen from the term proportional to $\rho_{ca}$. Although levels $\ket{a}$ and $\ket{c}$ are not directly coupled, the coherence created between them introduces a contribution to $\rho_{ba}$ due to a second field. 

This means that the susceptibility depends on additional parameters such as the Rabi frequency and dipole matrix elements from another transition, which affects its refractive index. Varying these parameters in general also changes $\rho_{ca}$, which can be prevented in a four-level system, as discussed in the next section.

Keeping linear dispersion in $\mathbf{E_1}$ but all orders of $\mathbf{E_2}$, the result for the relevant coherence is
\begin{equation}
    \rho_{ba} \;=\; \frac{\Omega_1}{2\frac{|\Omega_2|^2}{(i\gamma_0+\Delta_1-\Delta_2)} - (i(\Gamma_{1}+\Gamma_{2})+2\Delta_1)}.
\end{equation}
This depends on the second transition through $\Omega_2$, but not the amplitude of $\mathbf{E_1}$ or phases of either field. Varying $|\Omega_2|$ from zero to arbitrarily high would not result in any significant change in $n$, so this is still no better than the two-level result. At best, for small $\Gamma_2$ compared to $\Gamma_1$, this approaches the two-level result, but as $\Gamma_2$ becomes comparable to or greater than $\Gamma_1$, the maximum possible real part of the susceptibility becomes smaller than what is possible with two levels. This is shown in \fig{2_level_vs_3_level}c.

Since coherence effects in a three-level system do not help in enhancing the index, we will consider another effect, wave mixing, in the following section.

\section{Manipulating refraction using wave mixing}
\label{4L section}
\begin{figure}[ht] 
\centerline{
\includegraphics[width=.9\linewidth]{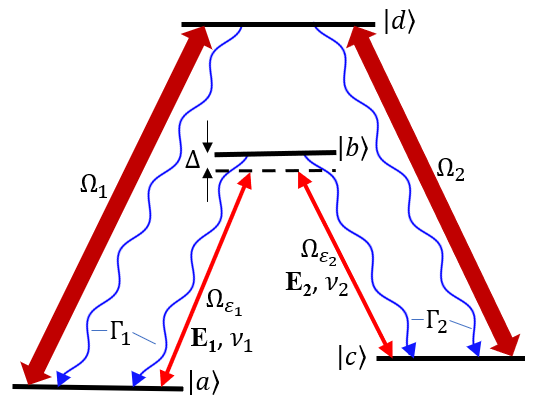}
}
\caption[Four-level system]{Four-level system. $\Omega_{\mathcal{E}_{1}}$, $\Omega_{\mathcal{E}_{2}}$, $\Omega_{1}$, and $\Omega_{2}$ are Rabi frequencies; $\Delta$ is a detuning, and $\Gamma_1$ and $\Gamma_2$ are decay rates. The $\ket{b}\rightarrow\ket{a}$ transition couples to the probe field $\mathbf{E_1}$ with angular frequency $\nu_1$, and the $\ket{b}-\ket{c}$ transition couples to the probe field $\mathbf{E_2}$ with angular frequency $\nu_2$.}
\label{4LA}
\end{figure}

In this section, wave mixing is considered in the context of enhancing the index. In the simplest case, wave mixing occurs with four levels, so we move to the system shown in \fig{4LA}. So far, the atomic and field properties seen in Section \ref{limits section} have not been useful in enhancing the index, except for the density, which has the practical limitations noted above. Field frequencies have appeared in detuning parameters, which cannot significantly enhance the index, but the frequencies may be more directly utilized. We show how direct frequency dependence is introduced with wave mixing, and calculate the index experienced by the probe field $\mathbf{E_1}$ as it depends on the frequency of another field $\mathbf{E_2}$.

The $\ket{d}\rightarrow\ket{a}$ and $\ket{d}\rightarrow\ket{c}$ transitions are strongly driven on resonance by external fields. The $\ket{b}\rightarrow\ket{a}$ electric dipole transition has moment $d_{1}$ (assumed real) and is coupled to the probe field $\mathbf{E_1}$ with complex amplitude $\mathcal{E}_1$ and frequency $\nu_1$, and the $\ket{b}\rightarrow\ket{c}$ transition has moment $d_{2}$ (assumed real) and is driven by the probe field $\mathbf{E_2}$ with amplitude $\mathcal{E}_2$ and frequency $\nu_2$.

For simplicity, we assume that the energy levels of $\ket{b}$ and $\ket{d}$ are similar with a transition frequency to $\ket{a}$ of $\omega_1$. The atomic transition frequency between $\ket{b}$ and $\ket{c}$ is $\omega_2$, which is approximately equal to the frequency between $\ket{d}$ and $\ket{c}$.

As seen in the previous section, the coherences that lead to the cross-coupling between the probe fields $\mathbf{E_1}$ and  $\mathbf{E_2}$ are $\rho_{ac}$ and $\rho_{ca}$. We want this cross-coupling to exist, while also having some way to enhance the linear part of the dispersion of a probe field. Here the fourth level allows for the additional strong driving fields to create and maintain $\rho_{ac}$ and $\rho_{ca}$, which in turn couple $\mathbf{E_1}$ and $\mathbf{E_2}$. We will show that changing $\omega_2$ while maintaining two-photon resonance on transitions $\ket{b}-\ket{c}$ and $\ket{d}-\ket{c}$ can affect the refractive index experienced by $\mathbf{E_1}$. This amounts to ``moving" level $c$ theoretically, which of course is not possible in practice, but this could be used to choose a level scheme in order to produce a potentially large refractive index.

There is a polarization due to the $\ket{b}\rightarrow\ket{a}$ transition and a polarization due to the $\ket{b}\rightarrow\ket{c}$ transition, which depend on the corresponding coherences, $\rho_{ba}$ and $\rho_{bc}$. $\rho_{ba}$ and $\rho_{bc}$ depend on the coherences $\rho_{ac}$ or $\rho_{ca}$ which are created by the strong fields. The polarization amplitudes to first order in $\mathcal{E}_1$ and $\mathcal{E}_2$ have the form
\begin{subequations}
\begin{align}
\mathcal{P}_1&\;=\;2Nd_{1}\rho_{ba}
&\;=\;\epsilon_0\chi_{11}\mathcal{E}_1+\epsilon_0\chi_{12}\mathcal{E}_2,
\label{P1}
\\
\mathcal{P}_2 &\;=\; 2Nd_{2}\rho_{bc}
&\;=\;\epsilon_0\chi_{21}\mathcal{E}_1+\epsilon_0\chi_{22}\mathcal{E}_2.
\label{P2}
\end{align}
\end{subequations}
From the Bloch equations, the four $\chi_{ij}$ terms can be found: 
\begin{equation}
\begin{split}
&\chi_{11} \;=\;2N\frac{d_{1}^2}{\hbar\epsilon_0}\alpha_{11}
\\
&\;=\;N\frac{6\pi c^3\Gamma_1}{\omega_1^3}\frac{|\Omega_{2}|^2}{[2\Delta-i(\Gamma_1+\Gamma_2)](|\Omega_{1}|^2+|\Omega_{2}|^2)},
\\
&\chi_{12} \;=\; 2N\frac{d_{1}d_{2}}{\hbar\epsilon_0}\alpha_{12}
\\
&\;=\; -N\frac{6\pi c^3\sqrt{\Gamma_1\Gamma_2}}{\sqrt{\omega_1^3\omega_2^3}}\frac{\Omega_{1}\Omega_{2}^*}{[2\Delta-i(\Gamma_1+\Gamma_2)](|\Omega_{1}|^2+|\Omega_{2}|^2)},
\\
&\chi_{21} \;=\; 2N\frac{d_{1}d_{2}}{\hbar\epsilon_0}\alpha_{21}
\\
&\;=\; -N\frac{6\pi c^3\sqrt{\Gamma_1\Gamma_2}}{\sqrt{\omega_1^3\omega_2^3}}\frac{\Omega_{1}^*\Omega_{2}}{[2\Delta-i(\Gamma_1+\Gamma_2)](|\Omega_{1}|^2+|\Omega_{2}|^2)},
\\
&\chi_{22} \;=\; 2N\frac{d_{2}^2}{\hbar\epsilon_0}\alpha_{22}
\\
&\;=\; N\frac{6\pi c^3\Gamma_2}{\omega_2^3}\frac{|\Omega_{1}|^2}{[2\Delta-i(\Gamma_1+\Gamma_2)](|\Omega_{1}|^2+|\Omega_{2}|^2)},
\end{split}
\label{chis}
\end{equation}
where $\chi_{12}$ corresponds to the $\ket{b}\rightarrow\ket{a}$ transition but comes from the cross-coupling with $\mathbf{E}_2$; it depends on the dipole moment of the $\ket{b}\rightarrow\ket{c}$ transition, which in turn depends on $\omega_2$; likewise, $\chi_{21}$ from the $\ket{b}\rightarrow\ket{c}$ transition depends on $\omega_1$. These do not depend on the amplitudes or phases of $\mathbf{E_1}$ or $\mathbf{E_2}$ since the polarizations are linear in the probe fields.

We assume that all fields are on resonance so that $\Delta=0$. This means that we must have $\nu_1=\omega_1$, and as $\omega_2$ is changed, the frequency of $\mathbf{E_2}$ must be chosen so that $\nu_2=\omega_2$. We also assume that for different positions of level $\ket{c}$, the decay rate $\Gamma_2$ remains roughly constant.

As with the two-level susceptibility, the $\chi_{ij}$ are proportional to the density and cannot be significantly increased by changing the Rabi frequencies, decay rates, or detuning. The important difference here is that $\mathcal{P}_1$ and $\mathcal{P}_2$ now depend on the frequencies of two atomic transitions, so the frequency corresponding to one transition can affect the susceptibility and polarization seen by the field coupled to the other transition, or equivalently, when the detuning is held constant, the overall susceptibility of one probe field depends on the frequency of the other probe field. Regardless of whether the detuning is held constant, which fixes the field frequencies in terms of the transition frequencies, direct dependence on the field frequencies is introduced when solving for the index experienced by a probe field for particular values of $\chi_{ij}$.

We now calculate the index of refraction experienced by $\mathbf{E_1}$ and see how it is affected by $\nu_2$. This is no longer as simple as using one susceptibility term in $n=\sqrt{1+\chi}$. The refractive indices that could possibly be attained in this system can be found by solving the Maxwell equations for the electric fields $\mathbf{E_1}$ and $\mathbf{E_2}$, which depend on the $\chi_{ij}$ found above. Maxwell's equations lead to equations for the amplitudes, $\mathcal{E}_1$ and $\mathcal{E}_2$, which depend on $z$:
\begin{subequations}
\begin{align}
\nabla^2 \mathcal{E}_1 &= -{\left( \frac{\nu_1}{c}\right)}^2 \left( 1+\chi_{11}\right) \mathcal{E}_1 - {\left( \frac{\nu_1}{c}\right)}^2 \chi_{12}\mathcal{E}_2, \\
\nabla^2 \mathcal{E}_2 &= -{\left( \frac{\nu_2}{c}\right)}^2 \chi_{21} \mathcal{E}_1 - {\left( \frac{\nu_2}{c}\right)}^2 \left( 1+\chi_{22}\right) \mathcal{E}_2.
\label{maxwell}
\end{align}
\end{subequations}
Finally, this shows the other way that the field frequencies appear. Using the ansatz $\mathcal{E}(z)=\mathcal{E}(0)e^{\lambda z}$, these equations can be solved, which results in two possible eigenvalues for $\lambda$. We are interested in the complex index seen by $\mathbf{E_1}$, which is related to an eigenvalue by $n=c\lambda/\nu_1$. The eigenvalues do not depend on the amplitudes and phases of either probe field, but the amplitudes and phases determine which eigenvalue and therefore which index value will be obtained. The corresponding eigenvectors give the initial amplitudes and phases of the probe fields that are required to obtain each eigenvalue. Any initial set-up for $\mathcal{E}_1$ and $\mathcal{E}_2$ is a combination of the two eigenvectors for that $\nu_2/\nu_1$. The part of $\mathcal{E}_1$ that projects onto one eigenvector experiences the index from the corresponding eigenvalue.

\begin{figure}[ht] 
\centerline{
\includegraphics[width=\linewidth]{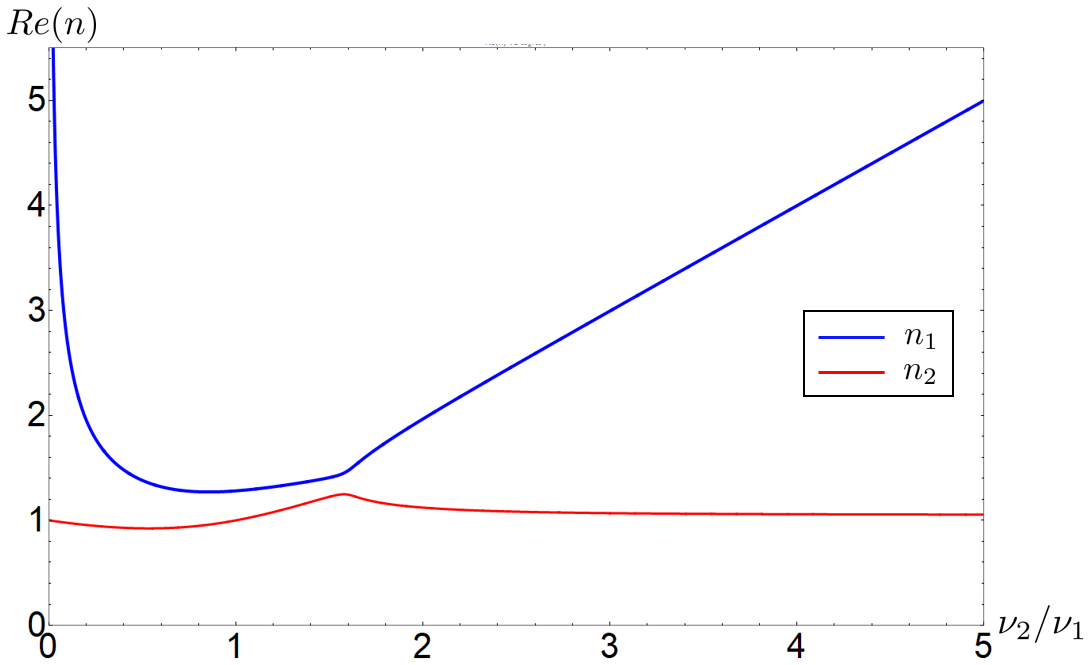}
}
\caption[Four-level index]{Real parts of the two possible eigenvalues for refractive index seen by $\mathbf{E_1}$ in the four-level system. $\Gamma_2=2\Gamma_1$, $\Omega_{1}=\Omega_{2}=\Gamma_1$, $\Delta=0$, $N=2.5\times10^{14}$ cm\textsuperscript{-3}, $\nu_1=\pi\times10^{15}$ Hz.}
\label{4_level_index_plot_re}
\end{figure}

\begin{figure}[ht] 
\centerline{
\includegraphics[width=\linewidth]{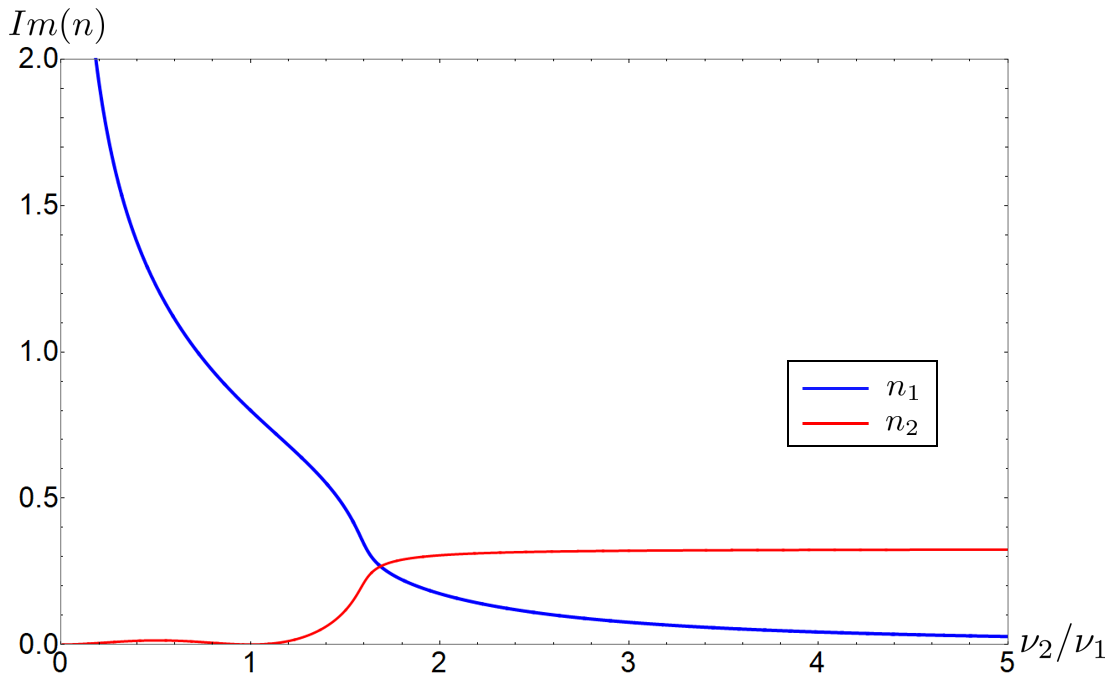}
}
\caption[Four-level index]{Imaginary parts of the two possible eigenvalues for refractive index seen by $\mathbf{E_1}$ in the four-level system. Parameters are the same as in \fig{4_level_index_plot_re}.}
\label{4_level_index_plot_im}
\end{figure}

The results show that the two possible indices for $\mathbf{E_1}$ are affected by the frequency of the other probe field, $\mathbf{E_2}$. This means that for the four-level system, there is a new effect which can be considered in order to obtain a desired refractive index. An example is shown in \fig{4_level_index_plot_re}, which shows the real parts of the possible indices for $\mathbf{E_1}$, which represent the amount of refraction experienced by the field. For $n_1$, the real part of the index shows some enhancement for $\nu_2/\nu_1<.5$ and for $\nu_2/\nu_1>1.6$. Incidentally, as shown in \fig{4_level_index_plot_im}, the absorption increases for smaller values of $\nu_2/\nu_1$, but decreases as $nu_2/\nu_1$ increases, suggesting that there may be a useful regime with an enhanced index and low absorption where $\nu_2/\nu_1$ increases beyond $1.6$.

\begin{figure}[ht]
\centerline{
\includegraphics[width=\linewidth]{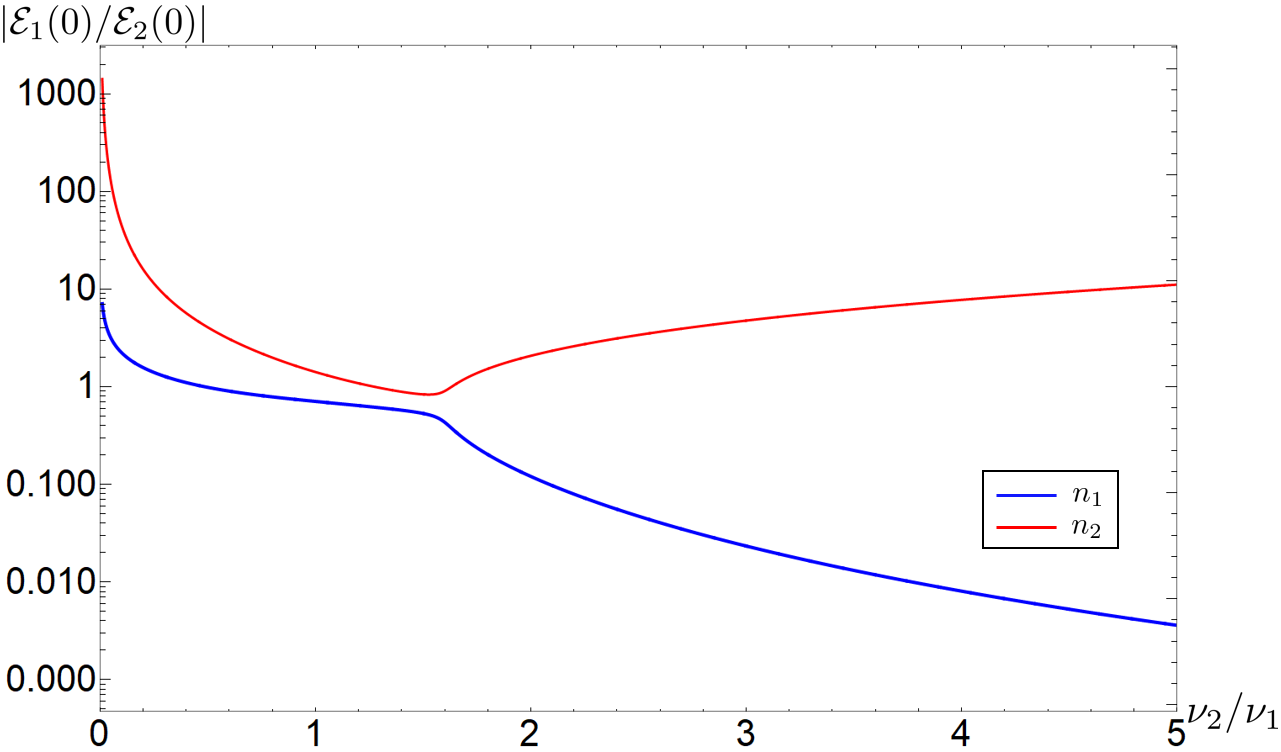}
}
\caption[Eigenvectors]{Plot on a logarithmic scale of the modulus of the eigenvector component for $\mathcal{E}_1$ relative to the eigenvector component for $\mathcal{E}_2$ for each possible eigenvalue as a function of $\nu_2/\nu_1$. Parameters are the same as in \fig{4_level_index_plot_re}.}
\label{4_level_index_eigenvector_amplitudes}
\end{figure}

The trade-off for this is seen in \fig{4_level_index_eigenvector_amplitudes}, which is a plot of the modulus of the eigenvector component corresponding to $\mathcal{E}_1$ relative to the modulus of the component corresponding to $\mathcal{E}_2$ for each eigenvalue. For small $\nu_2/\nu_1$ where $Re(n_1)$ is large, the required initial amplitude for $\mathcal{E}_2$ is comparable to or smaller than the required amplitude for $\mathcal{E}_1$. However, for larger $\nu_2/\nu_1$ where there is also an increase in $Re(n_1)$, the ratio of amplitudes becomes very small. For a given initial $\mathcal{E}_2$, less of $\mathcal{E}_1$ can be refracted at higher values of $\nu_2/\nu_1$, or else a much stronger $\mathcal{E}_2$ must be used to refract some amount of the field of interest. Otherwise, some of $\mathcal{E}_1$ will project onto the other eigenvector and experience $n_2$. Also, increasing $\nu_2$ (and simultaneously increasing $\omega_2$) decreases $\chi_{12}$, $\chi_{21}$, and $\chi_{22}$, which will eventually effectively decouple the fields. These observations suggest that there are limits to this behavior in the linear regime, in addition to the limitation imposed by the density. With wave mixing, there is more freedom in attaining a desired refractive index by choosing a particular level scheme, but this does not result in a significantly enhanced index in this simple case without nonlinear effects.


\section{Conclusion}
The only parameter that allows to change the amplitude over which the index of refraction can change in a given transition is, in principle, the density of the atomic gas. This is true for atomic gases and any other system. This, however, places severe limits on (three-dimensional, homogeneous) atomic gases due to disorder and dipole-dipole interactions \cite{chang21,Fleischhauer}, the total achievable index of refraction in an atomic gas seems limited to a small fraction of unity. Coherence effects, such as can be achieved in EIT-like systems, do not change this basic outcome.

Additional transitions and fields must be used to introduce other ways of modifying the refractive index. However, coherence effects at realistic densities do not help. To find some enhancement in the linear susceptibility, one needs to make use of the frequencies involved in the transitions.
This frequency dependence can be introduced via wave mixing in a more versatile many-level system (at least four) and allows for ways of producing a particular refractive index. Example results show how the resulting index experienced by one probe field on a lower transition is affected by the placement of the other lower level, but there are still limitations to this effect; larger enhancements require relatively smaller amplitudes of the field of interest.

Similar to the use of three-level systems above, these systems can now be expanded and optimized by coherence effects in order to make sure that, for example minimal absorption happens at the same detuning as maximal index of refraction. In addition, one can now include nonlinear effects if the goal is not just to manipulate the basic index of refraction but the effective dispersion of the medium. These will be studied in the future. 



\section{Acknowledgments}
We would like to acknowledge useful discussions with Marina Litinskaya and funding by the National Science Foundation via grants PHY-1912607 and PHY-1607637.

\section*{Appendix A: Two-Level Result Derivation}
\renewcommand{\thefigure}{A.\arabic{figure}}
\setcounter{figure}{0}
\renewcommand{\theequation}{A.\arabic{equation}}
\setcounter{equation}{0}
We go over the details of the derivation of the standard result shown in Sec. \ref{2L section} for completeness. Each two-level system has states $\ket{a}$ and $\ket{b}$ with atomic transition frequency $\omega$, and population decay rate $\Gamma$. There is an electric dipole transition with dipole operator $\hat{\mathbf{d}}$ between the levels; we neglect magnetic dipole transitions because they are typically weaker than electric dipole transitions and therefore would lead to a smaller magnetic susceptibility and a smaller index. A medium consists of $N$ of these two-level atoms, and we assume that there is no interaction among the atoms. The medium is driven by an electric field $\mathbf{E}$ with amplitude $\mathcal{E}$ and angular frequency $\nu$ as in Section \ref{Intro}, detuned from the two-level resonance by $\Delta$. The polarization is $\mathbf{P}=N\evalue{\hat{\mathbf{d}}} = N\hat{\mathbf{\epsilon}}(d\rho_{ab}+d^*\rho_{ba})$ \cite{text1, grynberg10, jackson98}, so we have
\begin{equation}
 \mathcal{P} = 2Nd^*\rho_{ba} = \epsilon_0\chi \mathcal{E}.
\label{pol}   
\end{equation}
Solving the steady-state Bloch equations under the dipole and rotating-wave approximations gives
\begin{equation}
  \rho_{ba}= \frac{i\Gamma-2\Delta}{\Gamma^2+4\Delta^2+2|\Omega|^2}\Omega.
\label{rho21}  
\end{equation}
To find $\chi$, we combine \eqs{pol} and \noeq{rho21}, and use the Rabi frequency in terms of the field amplitude: $\Omega=d \mathcal{E}/\hbar$. To replace $|d|^2$, we use the expression for spontaneous emission rate: $\Gamma = |d|^2 \omega^3 /(3\pi\hbar\epsilon_0 c^3) \approx 8\pi^2|d|^2/(3\hbar\epsilon_0 \lambda^3)$, where $\lambda$ is the wavelength of the incident light, and where we have assumed that $\Delta \equiv (\nu-\omega)\ll \nu, \omega$. The standard result is \cite{text1, Loudon}
\begin{equation}
  \begin{aligned}
\chi &=2N\frac{d^*\rho_{ba}}{\epsilon_0\mathcal{E}}
\\
&=N\frac{3\lambda^3\Gamma}{4\pi^2}\frac{i\Gamma-2\Delta}{\Gamma^2+4\Delta^2+2|\Omega|^2}.
\end{aligned}  
\end{equation}
The three-level derivation follows the same procedure but with $\mathbf{P}_1=N\evalue{\hat{\mathbf{d}}_1}=N\hat{\epsilon}_1(d_1\rho_{ab}+d^*_1\rho_{ba})$ and
$\mathbf{P}_2=N\evalue{\hat{\mathbf{d}}_2}=N\hat{\epsilon}_2(d_2\rho_{cb}+d^*_2\rho_{bc})$.

\section*{Appendix B: Plots of $\chi_{ij}$}
\renewcommand{\thefigure}{B.\arabic{figure}}
\setcounter{figure}{0}
\renewcommand{\theequation}{B.\arabic{equation}}
\setcounter{equation}{0}
Example plots of the $\chi_{ij}$ from \eq{chis} are shown in \fig{chi plots} for three values of $\nu_2=\omega_2$. For smaller $\nu_2$ as in \fig{chi_plots_0_1}, $\chi_{12}$, $\chi_{21}$, and $\chi_{22}$ increase overall. The $\nu_2^2$ factors in \eq{maxwell} decrease, so the $\chi_{21}$ term falls off, but the $\chi_{12}$ and $\chi_{22}$ terms increase overall which contributes to the index enhancement for small $\nu_2/\nu_1$. 

For larger $\nu_2$ as in \fig{chi_plots_10}, $\chi_{12}$, $\chi_{21}$, and $\chi_{22}$ decrease, and the $\chi_{12}$ and $\chi_{22}$ terms in \eq{maxwell} decrease, but the $\chi_{21}$ term increases which contributes to an index enhancement still for $\mathbf{E_1}$.

\begin{figure}[H]
\begin{subfigure}{\linewidth}
\caption[chi_plots]{$\nu_2=.1\nu_1$}
\includegraphics[width=\linewidth]{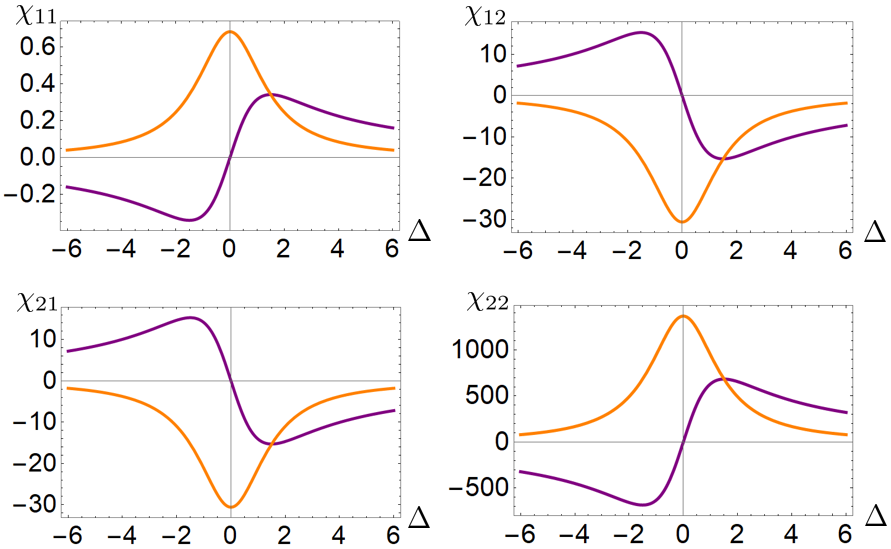}
\label{chi_plots_0_1}
\end{subfigure}
\begin{subfigure}{\linewidth}
\caption[chi_plots]{$\nu_2=\nu_1$}
\includegraphics[width=\linewidth]{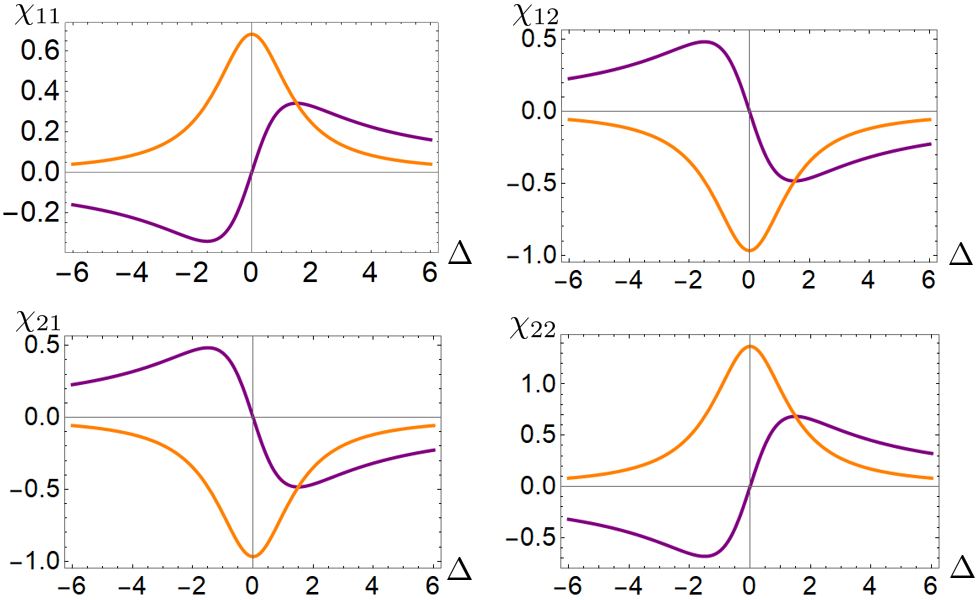}
\label{chi_plots_1}
\end{subfigure}
\begin{subfigure}{\linewidth}
\caption[chi_plots]{$\nu_2=10\nu_1$}
\includegraphics[width=\linewidth]{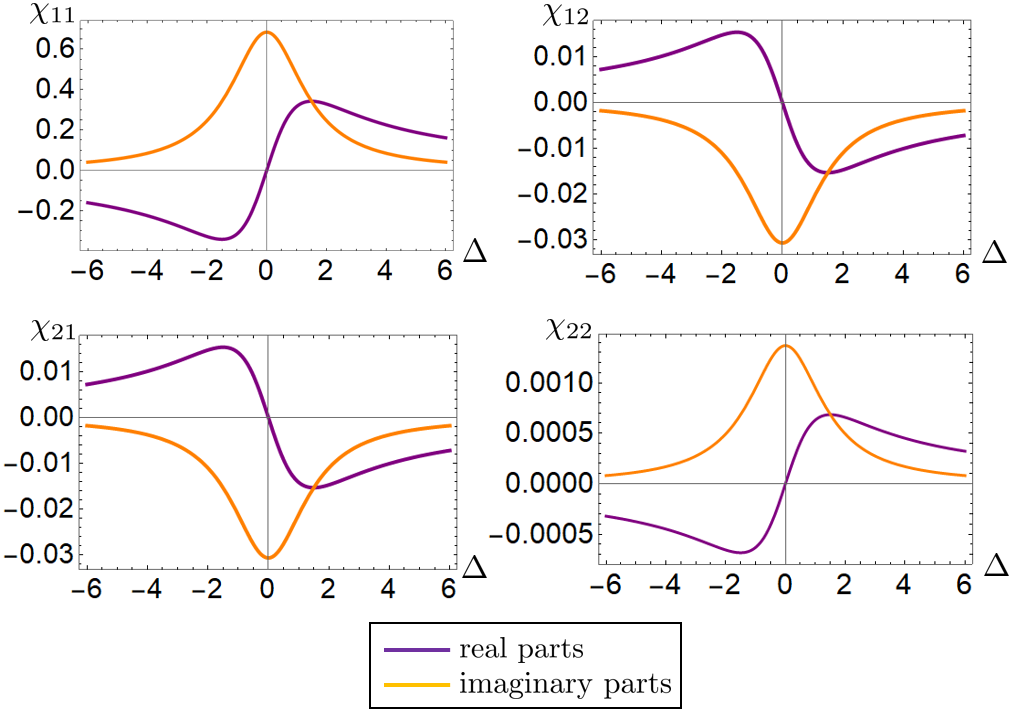}
\label{chi_plots_10}
\end{subfigure}
\caption{Plots of the four $\chi_{ij}$ with real parts (purple) and imaginary parts (orange) as functions of $\Delta$ for different values of $\nu_2$, with all other parameters the same as in \fig{4_level_index_plot_re}. Assuming $\nu_2=\omega_2$, increasing $\nu_2$ leads to an overall decrease in $\chi_{12}$, $\chi_{21}$, and $\chi_{22}$.}
\label{chi plots}
\end{figure}

\bibliographystyle{elsarticle-num}
\bibliography{indexlimit}

\begin{thebibliography}{10}
\expandafter\ifx\csname url\endcsname\relax
  \def\url#1{\texttt{#1}}\fi
\expandafter\ifx\csname urlprefix\endcsname\relax\def\urlprefix{URL }\fi
\expandafter\ifx\csname href\endcsname\relax
  \def\href#1#2{#2} \def\path#1{#1}\fi

\bibitem{Scully}
M.~O. Scully, Enhancement of the index of refraction via quantum coherence,
  Phys. Rev. Lett. 67~(14) (1991) 1855--1858.

\bibitem{Dowling-Bowden}
J.~P. Dowling, C.~M. Bowden, Near dipole-dipole effects in lasing without
  inversion: An enhancement of gain and absorptionless index of refraction,
  Opt. Express 70~(10) (1993) 1421--1424.

\bibitem{Y-F}
S.~F. Yelin, M.~Fleischhauer, Modification of local field effects in two level
  systems due to quantum corrections, Opt. Express 1~(6) (1997) 160--168.

\bibitem{F-Y}
M.~Fleischhauer, S.~F. Yelin, Radiative atom-atom interactions in optically
  dense media: Quantum corrections to the lorentz-lorenz formula, Phys. Rev. A
  59~(3) (1999) 2427--2441.

\bibitem{Fleischhauer}
M.~Fleischhauer, Electromagnetically induced transparency and coherent-state
  preparation in optically thick media, Opt. Express 4~(2) (1999) 107--112.

\bibitem{chang21}
F.~Andreoli, M.~J. Gullans, A.~A. High, A.~Browaeys, D.~E. Chang, Maximum
  refractive index of an atomic medium, Phys. Rev. X 11~(1) (2021) 011026.

\bibitem{meta1}
D.~R. Smith, W.~J. Padilla, D.~C. Vier, S.~C. Nemat-Nasser, S.~Schultz,
  Composite medium with simultaneously negative permeability and permittivity,
  Phys. Rev. Lett. 84~(18) (2000) 4184--4187.

\bibitem{meta2}
R.~A. Shelby, D.~R. Smith, S.~Schultz, Experimental verification of a negative
  index of refraction, Science 292 (2001) 77--79.

\bibitem{meta3}
S.~Linden, C.~Enkrich, M.~Wegener, J.~Zhou, T.~Koschny, C.~Soukoulis, Magnetic
  response of metamaterials at 100 terahertz, Science 306 (2005) 1351--1353.

\bibitem{Kastel}
J.~Kastel, M.~Fleischhauer, S.~F. Yelin, R.~L. Walsworth, Tunable negative
  refraction without absorption via electromagnetically induced chirality,
  Phys. Rev. Lett. 99~(99) (2007) 073602.

\bibitem{density1}
A.~P. Fang, W.~Ge, M.~Wang, F.~L. Li, M.~S. Zubairy, Negative refractive index
  without absorption via quantum coherence, Phys. Rev. A 93 (2016) 023822.

\bibitem{fieldcontrol}
X.~M. Su, H.~X. Kang, J.~Kou, X.~Z. Guo, J.~Y. Gao, Electromagnetically induced
  left-handedness by both coherent and incoherent fields, Phys. Rev. A 80
  (2009) 023805.

\bibitem{Marina}
M.~Litinskaya, E.~A. Shapiro, Negative refraction and photonic-crystal optics
  in a cold gas, Phys. Rev. A 91~(3) (2015) 033802.

\bibitem{cloaking}
J.~B. Pendry, D.~Schurig, D.~R. Smith, Controlling electromagnetic fields,
  Science 312 (2006) 1780--1782.

\bibitem{cloaking2}
U.~Leonhardt, Optical conformal mapping, Science 312 (2006) 1777--1780.

\bibitem{cloaking3}
T.~Ergin, N.~Stenger, P.~Brenner, J.~B. Pendry, M.~Wegener, Three-dimensional
  invisibility cloak at optical wavelengths, Science 328 (2010) 337--379.

\bibitem{cloaking4}
X.~Chen, Y.~Luo, J.~Zhang, K.~Jiang, J.~B. Pendry, S.~Zhang, Macroscopic
  invisibility cloaking of visible light, Nat. Comm. 2 (2010) 176.

\bibitem{ni}
V.~Veselago, The electrodynamics of substances with simultaneously negative
  values of epsilon and mu, Sov. Phys. Usp. 10 (1968) 509--514.

\bibitem{lens}
J.~B. Pendry, Negative refraction makes a perfect lens, Phys. Rev. Lett. 85
  (2000) 3966--3969.

\bibitem{nanooptics}
L.~Novotny, B.~Hecht, Principles of Nano-Optics, 1st Edition, Cambridge
  University Press, Cambridge, UK, 2006.

\bibitem{text1}
M.~O. Scully, M.~S. Zubairy, Quantum Optics, Cambridge University Press, New
  York, New York, 2008.

\bibitem{grynberg10}
G.~Grynberg, A.~Aspect, C.~Fabre, Introduction to Quantum Optics, Cambridge
  University Press, Cambridge, England, 2010.

\bibitem{jackson98}
J.~D. Jackson, Classical Electrodynamics, 3rd Edition, Wiley, New York, 1998.

\bibitem{Loudon}
R.~Loudon, The Quantum Theory of Light, Oxford University Press, New York,
  2000.

\bibitem{density4}
C.~Chin, A.~J. Kerman, V.~Vuletic, S.~Chu, Sensitive detection of cold cesium
  molecules formed on feshbach resonances, Phys. Rev. Lett. 90~(3) (2003)
  033201.

\bibitem{Thommen}
Q.~Thommen, P.~Mandel, Electromagnetically induced left handedness in optically
  excited four-level atomic media, Phys. Rev. Lett. 96 (2006) 053601.

\bibitem{density2}
P.~P. Orth, R.~Hennig, C.~H. Keitel, J.~Evers, Negative refraction with tunable
  absorption in an active dense gas of atoms, New J. Phys. 15 (2013) 013027.

\bibitem{density3}
Z.~Zhang, Z.~Liu, S.~Zhao, J.~Zheng, Y.~Ji, N.~Liu, Negative refractive index
  in a four-level atomic system, Chinese Phys. B 20 (2011) 124202.

\bibitem{javanainen16}
J.~Javanainen, J.~Ruostekoski, Light propagation beyond the mean-field theory
  of standard optics, Opt. Express 24~(2) (2016) 993--1001.

\bibitem{jennewein18}
S.~Jennewein, L.~Brossard, Y.~R.~P. Sortais, A.~Browaeys, P.~Cheinet,
  J.~Robert, P.~Pillet, Coherent scattering of near-resonant light by a dense,
  microscopic cloud of cold two-level atoms: Experiment versus theory, Phys.
  Rev. A 97~(5) (2021) 053816.

\bibitem{pellegrino14}
J.~Pellegrino, R.~Bourgain, S.~Jennewein, Y.~R.~P. Sortais, A.~Browaeys, S.~D.
  Jenkins, J.~Ruostekoski, Observation of suppression of light scattering
  induced by dipole-dipole interactions in a cold-atom ensemble, Phys. Rev.
  Lett 113~(13) (2014) 133602.

\bibitem{boyd92}
R.~W. Boyd, Nonlinear Optics, Academic Press, Boston, 1992.

\end{thebibliography}
\end{document}